# Identification of high energy ions using backscattered particles in laser-driven ion acceleration with cluster-gas targets

Y. Fukuda[a]*, H. Sakaki[a], M. Kanasaki[a,b], A. Yogo[a], S. Jinno[a], M. Tampo[a], A. Ya. Faenov[a,c], T. A. Pikuz[a,c], Y. Hayashi[a], M. Kando[a], A.S. Pirozhkov[a], T. Shimomura[a], H. Kiriyama[a], S. Kurashima[d], T. Kamiya[d], K. Oda[b], T. Yamauchi[b], K. Kondo[a], S. V. Bulanov[a]

[a] *Kansai Photon Science Institute (KPSI), Japan Atomic Energy Agency (JAEA),*
  *8-1-7 Umemidai, Kizugawa-shi, Kyoto 619-0215, Japan*
[b] *Graduate School of Maritime Sciences, Kobe University,*
  *5-1-1 Fukaeminami-machi, Higashinada-ku, Kobe 658-0022, Japan*
[c] *Joint Institute for High Temperatures, Russian Academy of Sciences, Moscow 125412, Russia*
[d] *Takasaki Advanced Radiation Research Institute, Japan Atomic Energy Agency,*
  *1233 Watanuki-machi, Takasaki-shi, Gunma 370-1292, Japan*

**Abstract**

A new diagnosis method for high energy ions utilizing a single CR-39 detector mounted on plastic plates is demonstrated to identify the presence of the high energy component beyond the CR-39's detection threshold limit. On irradiation of the CR-39 detector unit with a 25 MeV per nucleon He ion beam from conventional rf-accelerators, a large number of etch pits having elliptical opening shapes are observed on the rear surface of the CR-39. Detailed investigations reveal that these etch pits are created by heavy ions inelastically backscattered from the plastic plates. This ion detection method is applied to laser-driven ion acceleration experiments using cluster-gas targets, and ion signals with energies up to 50 MeV per nucleon are identified.

*Corresponding Author:
    e-mail: fukuda.yuji@jaea.go.jp

# 1. Introduction

The laser-driven ion acceleration via the interaction of short, intense laser pulses with matter, known as laser-plasma acceleration, is featured by its high accelerating electric fields and short pulse length compared to conventional rf-accelerators. It has been one of the most active areas of research during the last several years (Borghesi et al., 2006; Fuchs et al., 2006; Robson et al., 2007), because the laser-driven ion beams can be employed in a broad range of applications in cancer therapy (Bulanov et al., 2002; Habs et al., 2011), isotope preparation for medical applications (Spencer et al., 2001), proton radiography (Borghesi et al., 2002), and controlled thermonuclear fusion (Roth et al., 2001).

The recent advancements of novel ion acceleration techniques (Fukuda et al., 2009; Hening et al., 2009; Gaillard et al., 2011; Haberbergar et al., 2011; Willingale et al., 2011) have potentials to enhance the maximum energy of accelerated ions up to 80-200 MeV, the requirement for proton radiotherapy. For example, substantial enhancement of the accelerated ion energies up to 10-20 MeV per nucleon has been demonstrated by utilizing the unique properties of the cluster-gas target (Fukuda et al., 2009), corresponding to approximately tenfold increase in the ion energies compared to previous experiments using thin foil targets.

In the laser-driven ion acceleration experiments, however, contrary to experiments using conventional rf-accelerators, which provide a well characterized ion beam with a fixed energy, it is almost impossible to predict the maximum ion energy in advance, because of the nonlinear dependence of the maximum ion energy on various experimental conditions. Therefore, diagnosis of laser-accelerated ion beams attracts much attention. A precise characterization of accelerated ions is a crucial issue to develop a laser ion accelerator for medical and industrial applications. Solid state nuclear track detectors (Fleischer et al., 1975) such as CR-39 detectors have been so far extensively used to detect ions (Fukuda et al., 2009; Gaillard et al., 2007; Haberbergar et al., 2011; Tampo et al., 2010), because the CR-39 have a great advantage that they are insensitive to high energy photons and electrons and capable of detecting only ions (Oda et al., 1997). In principal, however, it is impossible to detect ions with energies higher than the detection threshold limit of the CR-39.

Therefore, new methods to detect ions whose energies are far beyond the detection threshold limit of the CR-39 are required to further facilitate the laser ion acceleration experiments. Here we demonstrate a new simple diagnosis method for high energy ions utilizing a single CR-39 detector mounted on plastic plates to identify the presence of the high energy component beyond the detection threshold limit of the CR-39. In addition, we have applied this method in combination with a magnetic energy spectrometer to the laser-driven ion acceleration experiments using cluster-gas targets to identify the higher energy component of accelerated ions.

## 2. Demonstration of the new diagnosis method for high energy ions
### 2-1. Experiment

Proof-of-principle experiments for the new ion detection method utilizing a single CR-39 detector mounted on plastic plates were carried out at Takasaki Ion Accelerators for Advanced Radiation Application (TIARA) facility in JAEA-TAKASAKI (Arakawa et al., 1992), which delivered 5-nA $^4He^{2+}$ ion beam with an energy of 100 MeV (25 MeV per nucleon). Figure 1(a) shows a side view of the ion detector unit, which consists of a 6-μm thick Al foil protector, 100-μm thick CR-39 (BRYOTRAK, Fukuvi Chemical Industry), a 3-mm thick PMMA, and a 2-mm thick Teflon. The rear surface of the CR-39 is directly contacted onto the plastic plate which works as a backscatterer and a moderator. The detector unit was exposed to the ion beam with a total fluence of $2.0 \times 10^8$ ions/cm$^2$. The irradiated CR-39 samples were chemically etched in a stirred 6M-KOH solution at 70 ℃ for total 6 hours using a multi-step etching technique (Oda et al., 1992).

### 2-2. Results and discussion

Although the energy of the incident He ion beam (25 MeV per nucleon) was far beyond the detection threshold limit (10 MeV per nucleon) of the CR-39 for He ions, i.e. the stopping power of the ion is too small to create etchable tracks inside the 100-μm CR-39, a number of etch pits were observed. Figures 1(b) and 1(c) show spatial distributions of the etch pits registered on the front and the rear surfaces of the CR-39, respectively, obtained by scanning the whole CR-39 surfaces by using a fast scanning microcopy (HSP1500, Seiko Precision, Co. Ltd.).

The surfaces of the etched sample were examined in greater details by using an optical microscope (BX60-F3, OLYMPUS). As shown in Figs. 2(a) and 2(b), we find that most of the etch pits on the rear surface have elliptical opening shapes, which indicates that high energy ions are obliquely injected into the CR-39 surface with random angles. On the other hand, almost no etch pit was observed when the single 100-μm thick CR-39 without the plastic plates was exposed to the ion beam. The sizes of the etch pit radii suggest that the pits are created by He and other heavy ions. These observations strongly suggest that the incident He ions (25 MeV per nucleon) once passed through the CR-39 without creating any etchable track, and that various nuclear reactions were induced by incident He ions in the plastic plates. As results, He and other heavy ions hit into the CR-39 from the rear surface, and the etchable tracks were created on the both rear and front surfaces via the processes (1) and (2) in Fig. 3. Note that the range for the incident 100-MeV (25 MeV per nucleon) He ions is calculated as 4.3 mm in the plastic plates.

This hypothesis was verified by investigating the growth curve behaviors of the etch pits using the multi-step etching technique (Kanasaki et al., 2011). Briefly, the results of the typical growth curve analyses for the three etch pits having elliptical opening shapes #1~#3 are shown in Fig. 2(c) and 2(d). The growth curve is defined as an evolution of a radius $r$ of a basal plane of a conical etch pit as a function of a thickness $G$ of a removed surface layer after the chemical etching process. Since the growth curves for $r$ and $r^2$ for the pits #1 and #2 are expressed as convex-downward increasing functions of $G$, respectively, we conclude that both the pits #1 and #2 are created by the backscattered ions. Similarly, since the growth curve for $r^2$ for the pit #3 is linearly proportional to $G$, while that for $r$ can be expressed as a convex-upward increasing function of $G$, we conclude that the pit #3 is also created by the backscattered ion.

The same analysis was carried out for 50 and 100 etch pits located inside the square boxes (see Figs. 1(b) and 1(c)) of the front and the rear surfaces, respectively. As a result, we conclude that almost all of the etch pits on the rear surface are created by the backscattered ions via the processes (1) and (2) in Fig. 3, and that on the front surface the 80 % of etch pits are created by the low energy ($E$ < 10 MeV per nucleon) background ions via the process (3) in Fig. 3, while the 20 % of etch pits are created by the backscattered ions that penetrated through from the rear surface via the process (2) in Fig. 3. The total number of the backscattered ions can be roughly estimated as $10^4$ ions/cm$^2$, which means that about 0.01 % of the incident ions are reflected back by the plastic plate.

Simulations using the PHITS code (Niita et al., 2010), which model the experimental configurations, have been performed to evaluate the yields of backscattered ions from PMMA/Teflon sheet to the rear surface of the CR-39 detector. The results show that 0.01~0.001 % of the incident ions are reflected back by the plastic plate. This is consistent with experimental result.

In summary, we demonstrated the new simple diagnosis method to identify the presence of the high energy ions beyond the CR-39's detection threshold limit using the backscattered particles from the plastic plates, where the plastic plates serve as a backscatterer.

## 3. Laser-driven ion acceleration experiments using cluster-gas targets

We have applied this ion detection method in combination with a magnetic energy spectrometer to the laser-driven ion acceleration experiments using the cluster-gas targets.

### 3-1. Experiment

The experiment has been conducted using the J-KAREN Ti:sapphire laser in JAEA-KPSI, the first petawatt-class OPCPA/Ti:sapphire hybrid laser system, which is designed to generate 37 fs pulses with a compressed energy of 20 J and a contrast ratio of $10^{-11}$ (Kiriyama et al., 2010). At the present experiment, we used 40-fs pulses with an energy of 1 J on target and a contrast ratio of $10^{-10}$. A schematic of experimental set up is shown in Fig. 4. The laser pulse was focused to a 30-μm diameter spot ($1/e^2$ intensity) with an off-axis parabolic mirrors with the effectivefocal lengths of 475 mm, which yields a peak intensity of $7\times10^{18}$ W/cm$^2$ in vacuum.

A pulsed solenoid valve connected to a specially designed circular nozzle (Boldarev et al., 2006) having a three-stage conical structure was used to produce submicron-size $CO_2$ clusters with an average diameter of ~0.4 μm (a root mean square deviation of ~0.1 μm) embedded in He gas. With the aid of a numerical model, we find that a 60-bar gas consisting of 90 % He and 10 % $CO_2$ is optimal for the production of submicron-size $CO_2$ clusters.

In order to identify high energy ions, we used a combination of a magnetic energy spectrometer (0.78 T, 100×100 mm) with a 16-mm entrance slit and the same ion detector unit as used in the poof-of-principle experiments, which consists of a 6-μm thick Al foil protector, a 100-μm thick CR-39 (BRYOTRAK, Fukuvi Chemical Industry), and a 3-mm thick PMMA, and a 2-mm thick Teflon with the size of 70×140 mm. The irradiated CR-39 sample was chemically etched in a stirred 6M-KOH solution at 70 ℃ for total 4.5 hours.

### 3-2. Results and discussion

Figure 5(a) shows the whole view of etch pit distribution on the front surface of the 70×140-mm wide CR-39, where red dots shows the etched pits including background noise. Ions were accumulated for 300 laser shots. The energies of ions registered on the CR-39 was calculated using 3D ion trajectory calculations assuming the observed ion signals to be fully stripped ions of $He^{2+}$, $C^{6+}$, and $O^{8+}$. Note that ions with the same energy-per-nucleon and the same atomic number to the nucleon number ratio of $Z/A = 1/2$, i.e., $He^{2+}$, $C^{6+}$, and $O^{8+}$ follow the same trajectory in the magnet. As shown in the inset of Fig. 4, depending on the initial direction of ion beams, the energy calculations give different values for the same ion energy. In other words, compared to the trajectory along the laser beam axis, the outer trajectory gives the lower energies, while the inner one gives the higher energies for the ion which was registered at the same position on the CR-39. Here we adopted results from the calculations for the outer trajectory so that the energy scale in Fig. 4 shows the "lower limit" for the accelerated ions.

As shown in Fig. 5(a), a large number of ions with energies ranging from 0.5 to 2 MeV per nucleon are observed. The cut-off around 0.5 MeV per nucleon well matches the stopping power of helium ions by the 6-μm Al foil. Note that helium, carbon, and oxygen ions with the energies exceeding 0.5, 0.8, and 0.9 MeV per nucleon, respectively, could reach the ion detector through the 6-μm thick Al foils. As shown in Fig. 5(b), detailed microscope investigations show that there exist at least two different sized etch pits in the same energy region. Since the track registration sensitivity depends strongly on ion mass (Dorschel et al., 2002) , we ascribe that the smaller pits are due to the helium ions and the larger ones to the carbon and/or oxygen ions. The ions observed on the front surface could be due to Coulomb explosion of $CO_2$ clusters and self-focusing channel.

According to the result shown in Fig. 5(a), no ions with energies greater than 2 MeV per nucleon were observed. However, as demonstrated in the previous section, for ions with energies greater than 10 MeV per nucleon, which is the detection threshold limit for the 100-μm thick CR-39, it is possible to create etchable tracks not on the front surface, but on the rear surface of the CR-39 by backscattered particles from the plastic plates. As shown in Fig. 5(c), we find that there exit a large number of etch pits on the rear surface. Detailed microscope investigations show that almost all of the etch pits on the rear surface have elliptical opening shapes (see Figs. 5(d) and 5(e)), which indicates that the high energy backscattered particles are obliquely injected into the CR-39 surface with random angles. Etch pit radii indicate that ion species are He and other heavy ions. Therefore, judging from analogy of the observation of the elliptical etch pits on the rear surface in the poof-of-principle experiments, we conclude that these elliptical etch pits are created by backscattered He and other heavy ions, i.e. laser-accelerated ions pass through the 100-μm CR39, and are backscattered by the plastic plates. As results, ions go back into the CR-39 and create etchable tracks on the rear surface of CR-39. As shown in Fig. 5(c), we could observe the elliptical etch pits up to 50 MeV per nucleon, which shows that ions with energy at least 50 MeV per nucleon are produced in laser-driven ion acceleration using the cluster-gas targets. Note that the etch pits on the rear surface are distributed in bunches, which indicates that the higher energy components of the ion beams have a mono-energetic nature. Details of acceleration mechanisms for this high energy ions will be discussed elsewhere.

Concerning the number of ions accelerated, since the poof-of-principle experiments showed that about 0.01 % of the incident ions were reflected back by the plastic plates, the number density of the ions is roughly estimated as $10^4$-$10^5$ cm$^{-2}$ per laser shot for 50 MeV per nucleon ions.

## 4. Conclusion

The proof-of-principle experiments for the new ion detection method were carried out at TIARA facility in JAEA, which delivered $^4$He$^{2+}$ ion beam with an energy of 100 MeV (25 MeV per nucleon). In contrast to the fact that the LET of the incident ion beam is too small to create etchable tracks in the 100-µm thick CR-39, a number of etch pits having elliptical opening shapes were created on the rear surface of the CR-39. The growth curve analyses using the multi-step etching technique revealed that the elliptical-shaped etch pits were created by the backscattered helium and other heavy ions, which were produced by various nuclear reactions in the plastic plates and hit into the CR-39 from the rear surface. About 0.01 % of the incident ions were reflected back by the plastic plates, which was supported by the PHITS code simulations.

This method in combination with a magnetic energy spectrometer was applied to the laser-driven ion acceleration experiments using the cluster-gas targets at J-KAREN laser facility in JAEA. On the rear surface, a number of etch pits having elliptical opening shapes were observed to demonstrate that laser-accelerated ions with energies up to 50 MeV per nucleon, which is greater than the detection threshold limit of the CR-39, passed through the CR39 without creating any etchable tracks, and that the backscattered ions hit into the CR-39 from the rear surface and created the etchable tracks on the rear surface which resulted in the etch pits having elliptical opening shapes.

**Acknowledgments**

We are grateful to Drs. T. Tajima, Y. Kishimoto, and T. Nakamura for valuable discussions. This work was supported by the Funding Program for Next Generation World-Leading Researchers from the Japan Society for the Promotion of Science.

**Figure Captions:**

Figure 1

(a) Schematic sectional view of the ion detector unit for the backscattering technique experiment, consisting of a 6-μm thick Al foil protector, 100-μm thick CR-39, a 3-mm thick PMMA, and a 2-mm thick Teflon. Spatial distributions of etch pits on (b) the front and (c) the rear surfaces of the CR-39. The red dot represents one etch pit. The growth curve analysis was carried out for the etch pits located inside the square boxes.

Fig. 2

(a), (b) Typical images of etch pits on the rear surface of the CR-39 detectors after etching for 6 hours. The growth curves for the etch pits #1~#3, plotting with (c) the semi-minor axis $r$ and (d) the squared one $r^2$ as a function of removed surface layer $G$.

Fig. 3

Four possible etch pit production processes (1)~(4).

Fig. 4

Schematic of the experimental setup for laser-driven ion acceleration experiments with the new ion detection method in combination with a magnetic energy spectrometer. The inset shows the top view of the ion energy analyzer, which consists of the 16-mm entrance slit, the 100×100-mm wide permanent magnet (0.78 T) and the 70×140-mm wide single CR-39 detector mounted on plastic plates. Compared to the trajectory along the laser beam axis, the outer trajectory (the blue line) gives the lower energies, while the inner one (the red line) gives the higher energies for the ion which was registered at the same position on the CR-39

Fig. 5

(a) Whole view of etch pit distribution on the front surface of the 70×140-mm wide CR-39. (b) Typical images of etch pits on the front surface. (c) Whole view of etch pit distribution on the rear surface of the 70×140-mm wide CR-39. (d), (e) Typical images of etch pits on the rear surface.

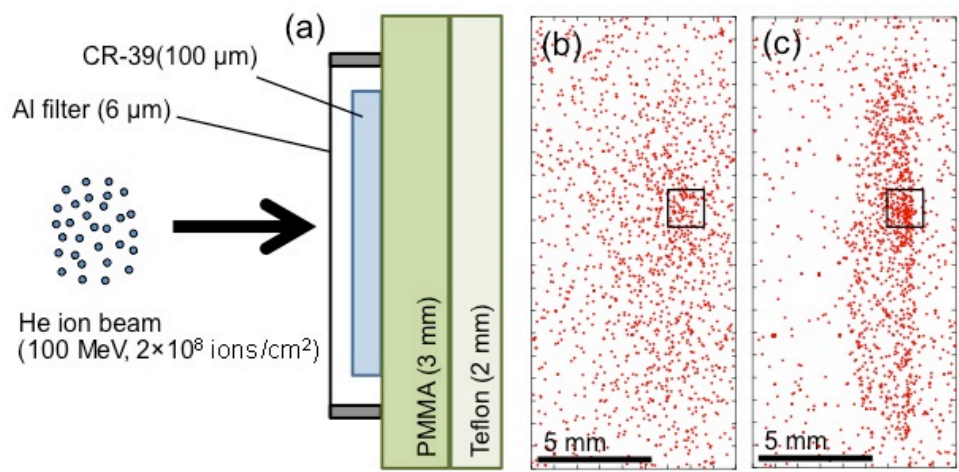

Figure 1. Y. Fukuda et al.

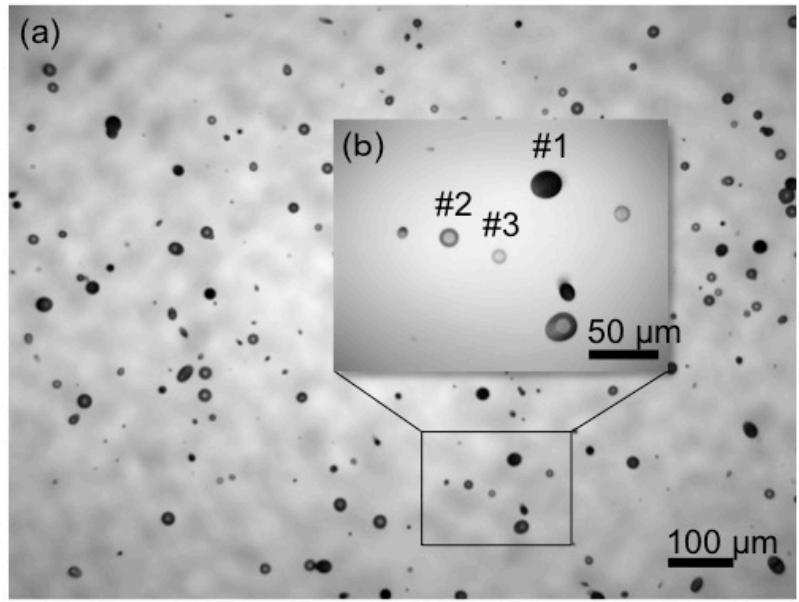
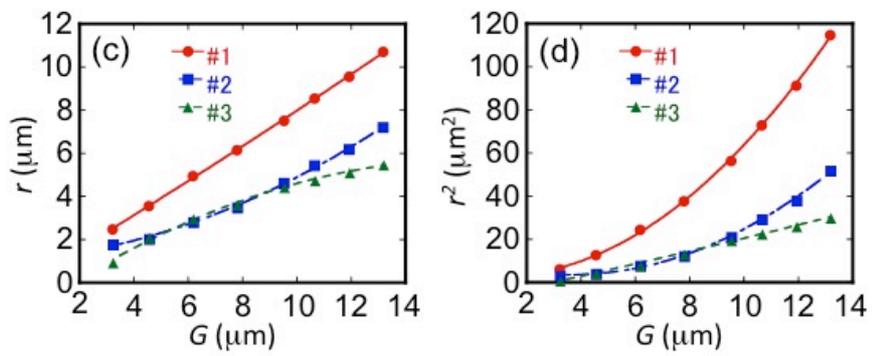

Figure 2. Y. Fukuda et al.

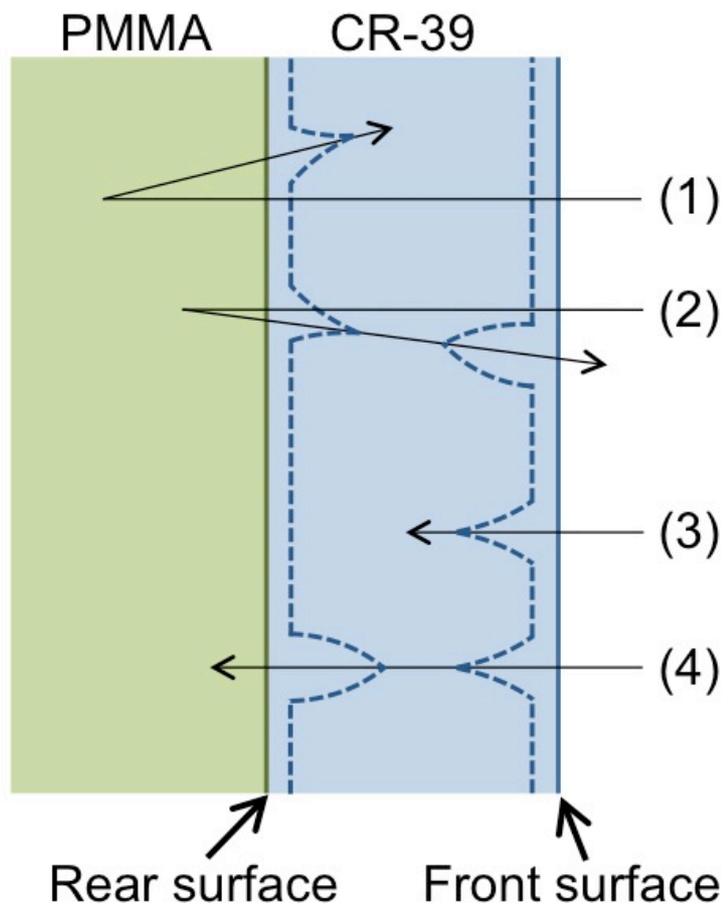

Figure 3. Y. Fukuda et al.

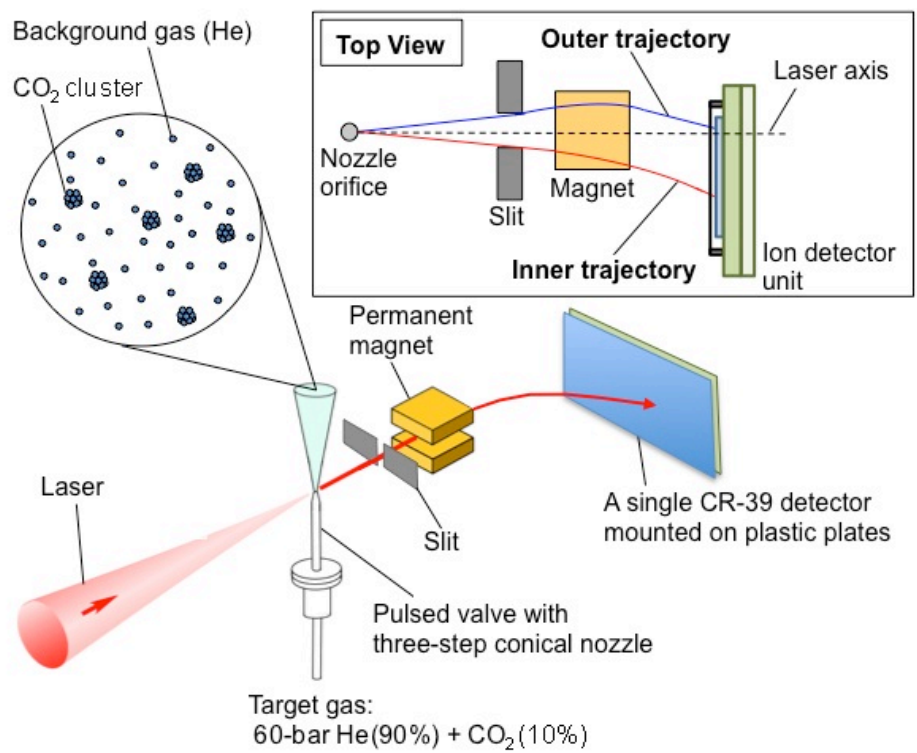

Figure 4. Y. Fukuda et al.

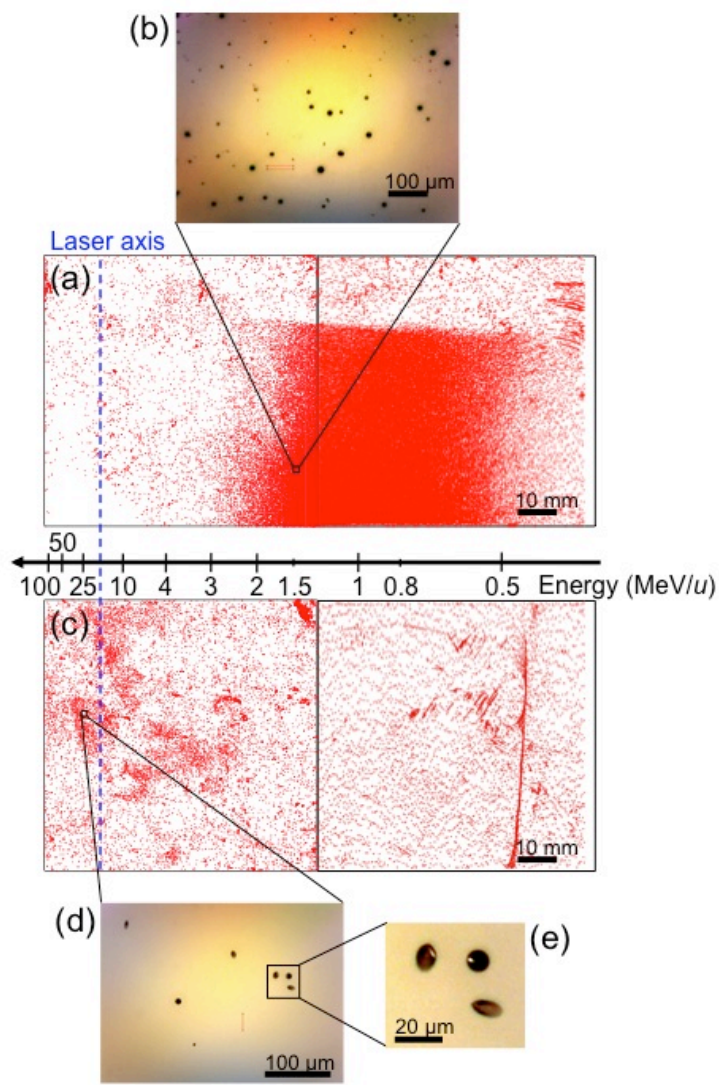

Figure 5. Y. Fukuda et al.